\documentclass[journal]{IEEEtran}
\usepackage{amsmath, amssymb, amsthm,mathtools, stmaryrd, bm, gensymb,cite}
\usepackage{xcolor}
\usepackage{caption,subcaption,siunitx}
\usepackage[nolist,nohyperlinks]{acronym}
\usepackage[ruled,vlined]{algorithm2e}
\usepackage{scalerel}
\usepackage{makecell}
\usepackage[none]{hyphenat}

\usepackage{url}
\usepackage{hyperref}

\newcommand{\bof}[1]{{\mbox{\boldmath$#1$}}}

\newcommand{\ys}[1]{\textcolor{black}{#1}}
\newcommand{\yss}[1]{\textcolor{black}{#1}}
\newcommand{\al}[1]{\textcolor{black}{#1}}
\newcommand{\fa}[1]{\textcolor{black}{{#1}}}

\begin{document}

\begin{acronym} % Give the longest label here so that the list is nicely aligned
    \acro{IRS}{intelligent reflecting surface}
    \acro{URA}{uniform rectangular array}
    \acro{BS}{base station}
    \acro{UE}{user-equipment}
    \acro{KF}{Kronecker factorization}
    \acro{LoS}{line-of-sight}
    \acro{SVD}{singular value decomposition}
    \acro{MIMO}{multiple input multiple output}
    \acro{SE}{spectral efficiency}
    \acro{SNR}{signal-to-noise ratio}
    \acro{B5G}{beyond fifth generation}
    \acro{EE}{energy efficiency}
    \acro{mmWave}{millimeter wave}
    \acro{MMSE}{minimum mean squared error}
    \acro{MISO}{multiple input single output}
    \acro{CSI}{channel state information}
    \acro{AoD}{azimuth of departure}
    \acro{EoD}{elevation of departure}
    \acro{AoA}{azimuth of arrival}
    \acro{EoA}{elevation of arrival}
    \acro{HOSVD}{high order singular value decomposition}
    \acro{TOT}{third-order tensor}
    \acro{AWGN}{additive white Gaussian noise}
    \acro{LSKRF}{least square Khatri-Rao factorization}
    \end{acronym}

	\bstctlcite{IEEEexample:BSTcontrol}
	\title{Low-Complexity Joint Active and Passive Beamforming Design for IRS-Assisted MIMO}
	\author{Yuri S. Ribeiro,  Fazal-E-Asim,~\IEEEmembership{Senior~Member,~IEEE}, André L. F. de Almeida,~\IEEEmembership{Senior~Member,~IEEE}, Behrooz Makki,~\IEEEmembership{Senior~Member,~IEEE}, and Gabor Fodor,~\IEEEmembership{Senior~Member,~IEEE}
				\thanks{This work was supported by the Ericsson Research, Sweden,
and Ericsson Innovation Center, Brazil, under UFC.51 Technical Cooperation Contract Ericsson/UFC. This study was financed in part by Funcap edital No. 05/2022. André L. F. de Almeida thanks CNPq for its financial support under grant 312491/2020-4. G. Fodor was partially
supported by the Digital Futures project PERCy. email: yurisales@gtel.ufc.br}}  % 
	\maketitle
	\begin{abstract}
		In this letter, we consider an intelligent reflecting surface (IRS)-assisted multiple input multiple output (MIMO) communication and we optimize the joint active and passive beamforming by exploiting the geometrical structure of the propagation channels. 
Due to the inherent Kronecker product structure of the channel matrix, the global beamforming optimization problem is split into lower dimensional horizontal and vertical sub-problems.
Based on this factorization property, we propose two closed-form methods for passive and active beamforming designs, at the IRS, the base station, and user equipment, respectively. The first solution is a singular value decomposition (SVD)-based algorithm independently applied on the factorized channels, while the second method resorts to a third-order rank-one tensor approximation along each domain. Simulation results show that exploiting the channel Kronecker structures yields a significant improvement in terms of computational complexity at the expense of negligible spectral efficiency (SE) loss. We also show that
under imperfect channel estimation, the tensor-based solution shows better SE than the benchmark and proposed SVD-based solutions.
  % \gf{\it Sanity check: each word in the title should appear or be expalined in the abstract, so I added "joint". However, what the abstract really emphasizes is low-complexity, not that much the spectral efficiency (where we actually loose a bit). So why not replace "Efficient" with "Low-complexity" in the title to make it consistent ?}
	\end{abstract}
	\section{Introduction}
	\label{sec:intro}
	\Ac{IRS} is a candidate technology to achieve high data rates required for \ac{B5G} wireless networks \cite{Basar2019,Gong2020}. 
 % Due to its low-cost structure, {IRSs can be easily deployed at existing \ac{BS}  sites or mounted on building facades and indoor walls} \cite{Gong2020}. 
The IRS is a two-dimensional planar array composed of multiple passive reflecting elements capable of changing the electromagnetic properties of the impinging waves, e.g., the phase and amplitude, so that the received signal can be added constructively at the receiver. On the other hand, \ys{terahertz (THz) communications suffer from high penetration and attenuation losses, which leads to sparse channels \cite{Andreas2023}}. To combat the path loss effects, passive beamforming using IRS may be introduced. Although it is an attractive solution for THz communications, the {joint} design of the active precoder, at the transmitter, combiner, at the receiver, and passive IRS phase shifts is challenging. 

In this regard, \cite{Zhang2018} proposed two joint passive and active beamforming methods {to maximize} the received signal power at the user in a \ac{MISO} network. 
Then, \cite{Song2021} and \cite{Rehman2021} provided two different approaches for joint active and passive beamforming in a multi-user \ac{MISO} networks.
% In \cite{Song2021}, the authors used a deep learning-based algorithm to maximize the sum rate of all users with low computational complexity. In \cite{Rehman2021}, the authors proposed an iterative vector approximate message passing-based solution to optimize the joint transmit beamforming and the \ac{IRS} phase shifts under the {sum minimum mean squared error} {criterion}. 
% This work further considers a more practical model for the IRS phase shifts.
% Although these works provide interesting results, they are focused on single antenna users. 
% By considering multiple antennas at both the base station and the users, \al{the complexity of the optimization problem becomes even higher when considering massive antenna arrays at both ends of the link}.
Regarding IRS-assisted \ac{MIMO}, \cite{zappone_overhead_aware} proposed three \ac{SVD}-based solutions for the joint active and passive beamforming design to maximize the \ac{SE} at the \ac{UE}. 
Also, \cite{Zhou2022} and \cite{Zhao2022} proposed an alternating optimization \al{scheme} to design the active and passive beamforming. 
% In \cite{Zhou2022}, the authors minimize \gf{the data detection mean squared error}, while in \cite{Zhao2022}, the authors optimize active and passive beamforming for different performance metrics.
In \cite{Bahingayi2022}, the authors proposed two low-complexity solutions for the beamforming design. 
Although some of those papers focus on low-complexity solutions, they still do not exploit the \al{explicit} geometrical structure of the channels to split the optimization problem into horizontal and vertical sub-problems with lower dimensions.  
% The first solution in \cite{zappone_overhead_aware} is the derivation of an upper bound, where the \ac{SVD} is applied in each channel to obtain both active and passive beamforming. The second in \cite{zappone_overhead_aware} is the derivation of a lower bound, where a single \ac{SVD} is applied on the cascaded channel to design joint active and passive beamforming. The third one in \cite{zappone_overhead_aware} is an alternating optimization approach, where the active beamforming is designed by fixing the passive beamforming and vice-versa until convergence.
% The proposed solutions become challenging either by increasing the number of transmit/receive antennas or the number of reflecting elements at the IRS. \\
\al{The works in \cite{Gil2021,Gil2022,Lin2021,asim2023,fazaleasim2023twodimensional} focused on channel estimation for IRS-assisted communications using tensor decomposition methods}, such as the {canonical polyadic decomposition. Furthermore, \cite{Sokal2022} and \cite{sokal2022reducing} proposed a rank-one tensor approximation to reduce the overhead associated with the IRS phase shift feedback.
To the best of our knowledge, there is no work that proposes tensor modeling for the joint active and passive beamforming design.

Our contributions can be summarized as follows:
\begin{enumerate}
    \item \al{We propose a novel signal modeling that exploits the geometrical channel structure at the \ac{BS}, the \ac{IRS}, and the \ac{UE} {by decomposing} the received signal into horizontal and vertical components. This approach allows splitting the joint active/passive \al{beamforming} problem into independent sub-problems {of} lower dimensions. Each sub-problem can be individually solved, \al{and} the solutions of \al{which} are combined using the Kronecker product to obtain the overall solution.}

    \item We propose two algorithms that exploit the \al{Kronecker product structure of the cascaded MIMO channel} to design the active and passive beamforming. The first method, \al{referred to as \ac{KF}, directly exploits the Kronecker structure by means of \ac{SVD}-based rank-one approximations applied on the factorized channels. Our second solution, referred to as \ac{TOT}, recasts the cascaded channel along each domain as a third-order rank-one tensor and resorts to the \ac{HOSVD} algorithm to optimize the active and passive beamforming vectors.}
    % \textcolor{blue}{Write about complexity comparison and some performance analysis}
\end{enumerate}
\ys{We show that our proposed TOT and KF solutions can reduce the computational complexity, respectively, by $15$ and $140$ times, compared to the benchmark scheme of \cite{zappone_overhead_aware} with similar \ac{SE} performance under perfect \ac{CSI} assumption. \yss{Otherwise, when imperfect \ac{CSI} is considered  the TOT method shows a significant SE improvement compared to the KF and the benchmark \cite{zappone_overhead_aware} methods.}}
% \textcolor{black}{but with $60$ and $10$ times less computational complexity with the KF and the \ac{TOT} methods, respectively. }

% \textbf{Notations}: Scalars are represented by lowercase letters ($a$), vectors are represented by boldface lowercase italic letters ($\bof{b}$), matrices are represented by boldface uppercase italic letters ($\bof{C}$) and tensors are represented by uppercase calligraphy letters ($\mathcal{D}$). The superscripts $\{\}^T$, $\{\}^*$ and $\{\}^H$ represent transpose, conjugate and hermitian operations, respectively. $\text{vec}(\bof{E})$ converts $\bof{E}$ to a column vector by stacking its columns on top of each other. The operators $\otimes$, $\diamond$ and $\odot$ represent the Kronecker, Khatri-Rao and Haddamard products, respectively.

%\vspace{-1ex}
%\noindent 
\textit{Notation:} \textcolor{black}{Scalars, vector, matrices and tensors are denoted ($a$),  ($\bof{a}$),  ($\bof{A}$) and ($\mathcal{A}$), respectively.} The superscripts $\{\}^T$, $\{\}^*$, and $\{\}^H$ denote transpose, conjugate, and hermitian, respectively. The operators $\otimes$, $\diamond$, $\circ$, $\odot$ and \textcolor{black}{$\angle$} are the Kronecker, the Khatri-Rao, the outer product, the Hadamard products and the \textcolor{black}{angle of a complex value}, respectively. %$\R{diag}[\bof{a}]$ is a  diagonal matrix holding the vector $\bof{a}$ as its main diagonal. 
$\text{vec}(\bof{A})$ converts $\bof{A}$ to a column vector by stacking its columns. %A third-order tensor $\mathcal{Y} \in \mathbb{C}^{I \times J \times K}$ can be unfolded as three different matrices, referred to as the $1$-mode, $2$-mode, and $3$-mode unfoldings, respectively, where $[\mathcal{Y}]_{(1)} \in \mathbb{C}^{I \times JK}$, $[\mathcal{Y}]_{(2)} \in \mathbb{C}^{J \times IK}$, and $[\mathcal{Y}]_{(3)} \in \mathbb{C}^{K \times IJ}$. 
The $n$-mode product of a tensor $\mathcal{X} \in \mathbb{C}^{I \times J \times K}$ and a matrix $\bof{A} \in \mathbb{C}^{I \times R}$ is denoted by $\mathcal{Y}=\mathcal{X}\times_n\bof{A}$, $n=1,2,3$. %For example, $[\mathcal{Y}]_{(1)} = \bof{A}[\mathcal{X}]_{(1)} \in \mathbb{C}^{R \times JK}$ is the result of the $1$-mode product between the tensor $\mathcal{Y}$ and the matrix $\bof{A}$.

% In this paper, we are exploiting the inherent structure of the geometrical channel to split the whole problem into horizontal and vertical domains with lower dimensions, which ultimately helps the design of both active and passive beamforming. 
% 	\textcolor{blue}{Write contributions of the paper, for instance, how many algorithms are proposed, etc.}

\section{System Model}
	\label{sec:sysmodel}
We consider a \ac{MIMO}-IRS communication system, where the \ac{BS} has $M$ antennas transmitting \ys{a single stream symbol} towards a \ac{UE} with $K$ antennas \textit{via} an \ac{IRS} having $N$ reflecting elements, which are adjusted by a controller connected to the \ac{BS}. \ys{Initially, Perfect \ac{CSI} of all channels is assumed at the \ac{BS} (see Section \ref{sec:simulations} for the effect of imperfect CSI)}. We further assume there is no direct link between the \ac{BS} and the \ac{UE}. 
% \gf{\it GF: Does the 'CSI' here refer to the channel between the BS and the IRS or to the channel between the BS and the UE or also to the channel between the IRS and the UE?}
The received signal $r$ at the \ac{UE} can be written as
\begin{align}
    &r = \bof{w}^H \bof{G}\text{diag}(\bof{\theta}) \bof{H} \bof{q} x + n,
    \label{eq:signal_model}
\end{align}
where $\bof{w} \in \mathbb{C}^{K\times1}$ is the combiner, $\bof{G} \in \mathbb{C}^{K\times N}$ is the channel between the \ac{IRS} and the UE, $\bof{\theta} \in \mathbb{C}^{N}$ is the \ac{IRS}
phase shift vector defined as $\bof{\theta} \doteq [e^{j\theta_1},\dots,e^{j\theta_N}]^T$, 
where $\theta_n \in [-\pi,\pi]$ represents the phase shift of the $n$-th reflecting element,  $\bof{H} \in 
\mathbb{C}^{N\times M}$ is the channel between the BS and the IRS, $\bof{q} \in 
\mathbb{C}^{M\times 1}$ is the transmit precoder, $x$ is the transmitted signal, with power $P_t$, and $n \sim \mathcal{C}\mathcal{N}
(0,\sigma_n^2)$ is the \ac{AWGN} with zero mean and variance $\sigma_n^2$.
% \gf{\it GF: Maybe we should state explicitly that in this paper, we consider a single stream transmission. This is not trivial, since in the Intro we emphasize that our model allows for multiple antennas at the UE.} 

\al{Our goal is to maximize the {\ac{SNR} at the \ac{UE}} subject to the \ac{IRS} phase shifts, precoder, and combiner constraints, which leads to the following optimization problem
\begin{align}
    &\max_{\bof{w},\bof{q},\bof{\theta}} |(\bof{w}^H \bof{G} \text{diag}(\bof{\theta}) \bof{H} \bof{q})|^2
    \label{eq:optimization_problem}\\
    &\text{s.t.} \ ||\bof{w}|| = ||\bof{q}|| = 1 \ \text{and} \  \theta_n \in [-\pi,\pi]. \nonumber
   % \label{eq:constrains}
\end{align}
A classical solution to this problem was proposed in \cite{zappone_overhead_aware} and related works, which  relies on \ac{SVD}-based steps applied to the full MIMO channel matrices $\bof{H}$ and $\bof{G}$ to find their dominant eigenmodes, from which the active beamforming vectors (precoder $\bof{q}$ and combiner $\bof{w}$), as well as the IRS phase shift vector $\bof{\theta}$ are determined.
}
\vspace{-0.1in}
\al{\section{Proposed approaches}}
\al{Here, we discuss our proposed low-complexity active and passive beamforming design. We first derive a Kronecker-based model for the involved channels and then recast the resulting two-dimensional optimization problem as a product of two smaller one-dimensional problems for each channel dimension. Then, our two algorithms exploiting the resulting optimization problem are derived.}

%\vspace{-1ex}
\al{\subsection{Kronecker-structured Channel Factorization}}
A \ac{URA} is deployed both at the \ac{BS} and the \ac{UE}, where the BS is \al{equipped with} $M_y$ antenna elements along the $y$ axis and $M_z$ antenna elements along the $z$ axis, \al{with $M= M_y M_z$ being the total number of antenna elements.} Similarly, the UE \al{has $K= K_y K_z$ antenna elements, where $K_y$ and $K_z$ are the number of elements along the $y$ axis and $z$ axis, respectively. \textcolor{black}{The element spacing between antennas of the URAs is $\lambda/2$.} At the BS side,} the response of the $m$th antenna element is defined as \cite{Fazal2021}
\begin{align}\label{eq:bs_steering}
    &[\bof{a}(\phi_{\text{bs}},\theta_\text{bs})]_m = e^{-j\pi[m_y\sin{\theta_\text{bs}} \sin{\phi_\text{bs}} + m_z\cos\theta_\text{bs}]},
\end{align}
where $m = m_z + (m_y - 1)M_z$ with $m_y~\in~{[0,1,\dots,M_y - 1]}$ and $m_z~\in~{ [0,1,\dots,M_z-1]}$, $\theta_\text{bs}$ and $\phi_\text{bs}$ represent the \ac{EoD} and \ac{AoD}, respectively. {Furthermore, define} the spatial frequencies as  $\mu_\text{bs} = \pi\sin\theta_\text{bs} \sin\phi_\text{bs}$ and $\psi_\text{bs} = \pi \cos\theta_\text{bs}$. The overall \al{array response in (\ref{eq:bs_steering})} can be written as a Kronecker product between the \al{associated} horizontal and the vertical \al{components} as $\bof{a}(\mu_\text{bs},\psi_\text{bs}) = \bof{a}_{y}(\mu_\text{bs}) \otimes \bof{a}_z(\psi_\text{bs}) \in \mathbb{C}^{M\times 1}$, where $\bof{a}_y(\mu_\text{bs}) $ is the horizontal steering vector as
    $\bof{a}_y(\mu_\text{bs}) = [1,e^{-j\mu_\text{bs}},\dots,e^{-j(M_y-1)\mu_\text{bs}}] \in \mathbb{C}^{M_y\times 1},$
and $\bof{a}_z(\psi_\text{bs})$ is the vertical  steering vector as $
    \bof{a}_z(\psi_\text{bs}) = [1,e^{-j\psi_\text{bs}},\dots,e^{-j(M_z - 1)\psi_\text{bs}}] \in \mathbb{C}^{M_z\times 1}.$
\al{Note that in a similar way,} the \ac{IRS} arrival, the \ac{UE}, and \ac{IRS} departure channel steering vectors can be written as 
$\bof{b}(\mu_{\text{irs}_A},\psi_{\text{irs}_A}) = \bof{b}_y(\mu_{\text{irs}_A})\otimes \bof{b}_z(\psi_{\text{irs}_A})$, $\bof{c}(\mu_{\text{ue}},\psi_{\text{ue}}) = \bof{c}_y(\mu_{\text{ue}})\otimes \bof{c}_z(\psi_{\text{ue}})$ and $\bof{d}(\mu_{\text{irs}_D},\psi_{\text{irs}_D}) = \bof{d}_y(\mu_{\text{irs}_D})\otimes \bof{d}_z(\psi_{\text{irs}_D})$, respectively.

\al{Under the previous definitions and assumptions, the channel matrix $\bof{H}$ linking the BS to the IRS can be written as}
\vspace{-1.5ex}
\begin{align}
    % &\bof{H} = \sum^R_{r=1}\alpha^{(r)}_H [\bof{s}(\mu_{\text{irs}_A}^{(r)},\psi_{\text{irs}_A}^{(r)})] [\bof{s}(\mu_\text{bs}^{(r)},\psi_\text{bs}^{(r)})]^T,\\
    &\hspace{-1.5ex}\bof{H}\!=\! \sum^L_{l=1}\alpha^{(l)}_H [\bof{b}_y(\mu^{(l)}_{\text{irs}_A}) \!\otimes\! \bof{b}_z(\psi^{(l)}_{\text{irs}_A})][\bof{a}_y(\mu^{(l)}_\text{bs}) \!\otimes\! \bof{a}_z(\psi^{(l)}_\text{bs})]^T\label{eq:H channel1}
\end{align}

\noindent where $L$ is the number of paths, and $\alpha^{(l)}_H$ $~\sim~$$\mathcal{CN}~(\mu,\sigma)$ is the complex gain of the $l$th path. 
% \gf{\it GF: It is a bit unfortunate that we also denoted the received signal with $r$, which is also the index of the multipath.}
Applying the property $(\bof{A}\otimes\bof{B})(\bof{C}\otimes\bof{D}) = \bof{A}\bof{C}\otimes \bof{B}\bof{D}$, we can rewrite \eqref{eq:H channel1} as
\begin{align}
&\bof{H}= \sum^L_{l=1}\underbrace{\alpha^{(l)}_H [\bof{b}_y(\mu^{(l)}_{\text{irs}_A}) \bof{a}^T_y(\mu^{(l)}_\text{bs})]}_{\bof{H}^{(l)}_y \in \mathbb{C}^{N_y\times M_y}}\otimes\underbrace{[\bof{b}_z(\psi^{(l)}_{\text{irs}_A}) \bof{a}^T_z(\psi^{(l)}_\text{bs})]}_{\bof{H}^{(l)}_z \in \mathbb{C}^{N_z \times M_z}} \label{eq:H channel2},
\end{align}
\al{or, compactly,}
\vspace{-3ex}
\begin{align}
    &\bof{H} = \sum^L_{l=1}\bof{H}^{(l)}_y \otimes \bof{H}^{(l)}_z.
    \label{eq:H channel}
\end{align}
% where $\bof{H}_y^{(r)} \in \mathbb{C}^{N_y\times M_y}$ is defined as $\alpha^{(r)}_H \bof{h}_y(\mu^{(r)}_{irs_A}) \bof{h}^T_y(\psi^{(r)}_{bs})$ and $\bof{H}^{(r)}_z \in \mathbb{C}^{N_z \times M_z}$ is defined as $\bof{h}_z(\mu^{(r)}_{irs_A}) \bof{h}^T_z(\psi^{(r)}_{bs})$.
\al{Similarly, following the same construction and definitions in (\ref{eq:H channel1}) and (\ref{eq:H channel2}) the IRS-UE channel $\bof{G}$ can be written as}
\vspace{-1.5ex}
\begin{align}
     &\al{\bof{G} = \sum^R_{r=1}\underbrace{\alpha^{(l)}_G [\bof{c}_y(\mu^{(l)}_{\text{ue}}) \bof{d}^T_y(\mu^{(l)}_{\text{irs}_D})]}_{\bof{G}_y^{(l)} \in \mathbb{C}^{K_y \times N_y}}\otimes\underbrace{[\bof{c}_z(\psi^{(l)}_{\text{ue}}) \bof{d}^T_z(\psi^{(l)}_{\text{irs}_D})]}_{\bof{G}_z^{(l)} \in \mathbb{C}^{K_z \times N_z}},}
    \label{eq:G channel1}
\end{align}
% \begin{align}
%     &\bof{G} = \sum^R_{r=1}\underbrace{\alpha^{(r)}_G [\bof{s}_y(\mu^{(r)}_{\text{ue}}) \bof{s}^T_y(\mu^{(r)}_{\text{irs}_D})]}_{\bof{G}_y^{(r)} \in \mathbb{C}^{K_y \times N_y}}\otimes\underbrace{[\bof{s}_z(\psi^{(r)}_{\text{ue}}) \bof{s}^T_z(\psi^{(r)}_{\text{irs}_D})]}_{\bof{G}_z^{(r)} \in \mathbb{C}^{K_z \times N_z}}
% \end{align}
\ys{and, analogously as in \eqref{eq:H channel}
\vspace{-1.5ex}
\begin{align}
    &\bof{G}  = \sum^L_{l=1} \bof{G}^{(l)}_y \otimes \bof{G}^{(l)}_z.
    \label{eq:G channel}
\end{align}}
% \al{where $\bof{G}^{(l)}_y\hspace{-1ex}=\hspace{-.5ex}\alpha^{(l)}_G \bof{s}_y(\mu^{(l)}_{\text{ue}}) \bof{s}^T_y(\mu^{(l)}_{\text{irs}_D})$, $\bof{G}^{(l)}_z\hspace{-1ex}=\hspace{-.5ex}\bof{s}_z(\psi^{(l)}_{\text{ue}}) \bof{s}^T_z(\psi^{(l)}_{\text{irs}_D})$,  $\bof{s}_y(\mu^{(l)}_{\text{ue}})$ and $\bof{s}_z(\psi^{(l)}_{\text{ue}})$ are respectively the $y$ and $z$ domain arrival steering vectors at the UE, while $\bof{s}_y(\mu^{(l)}_{\text{irs}_D})$ and $\bof{s}_z(\psi^{(l)}_{\text{ue}})$ are the $y$ and $z$ domain departure steering vectors at the IRS, respectively, associated with the $l$th path, $l=1,\ldots, L$.}
% are defined analogously as $\bof{H}^{(l)}_y$ and $\bof{H}^{(l)}_z$ in \eqref{eq:H channel2} 

\al{In practical scenarios, the channel {shows} less variation} in the vertical domain {compared to} the horizontal domain \cite{Hanzo2017}. \al{For example, in THz communications the channels are dominated by their \ac{LoS} components, while some non-LoS scenarios exhibit small angular spreads \cite{Andreas2023}.
In such scenarios, the vertical spatial frequencies of both $\bof{H}$ and $\bof{G}$ channels, defined in (\ref{eq:H channel}) and (\ref{eq:G channel}), are strongly correlated, which implies reasonably assuming that $\bof{H}^{(1)}_z \approx \cdots \approx \bof{H}^{(L)}_z$, and $\bof{G}^{(1)}_z \approx \cdots \approx \bof{G}^{(L)}_z$. This leads to the following approximations}
% $\psi^{(1)}_\text{bs} \approx \dots \approx \psi_{\text{bs}}^{(r)}, \ \psi^{(1)}_{\text{irs}_A} \approx \dots \approx \psi^{(r)}_{\text{irs}_A}, \
%   \psi^{(1)}_{\text{irs}_D}\approx \dots \approx \psi^{(r)}_{\text{irs}_D}, \ \psi^{(1)}_\text{ue} \approx \dots \approx \psi^{(r)}_\text{ue}$. 
% Therefore, the channels described in \eqref{eq:H channel} and \eqref{eq:G channel} can be approximately written as:
\begin{align}
    &\bof{H} \approx \underbrace{\Big(\sum^L_{l=1} \bof{H}^{(l)}_y\Big)}_{\bof{H}_y} \otimes \ \bof{H}_z = \bof{H}_y\otimes \bof{H}_z, \label{eq:H channel approx}\\
        &\bof{G} \approx \underbrace{\Big(\sum^L_{l=1} \bof{G}^{(l)}_y\Big)}_{\bof{G}_y} \otimes \ \bof{G}_z = \bof{G}_y\otimes \bof{G}_z.
    \label{eq:G channel approx}
\end{align}
\al{Otherwise stated, the involved MIMO channel matrices can be factorized in terms of the Kronecker product of their respective horizontal ($y$-domain) and vertical ($z$-domain) components. Note that such a Kronecker-structured model is exact in a pure  LoS scenario, or {when} the non-LoS components are negligible. }

\al{\subsection{Two-dimensional Active and Passive Beamforming}}
\al{By adopting the Kronecker-product based MIMO channel factorizations in (\ref{eq:H channel approx}) and (\ref{eq:G channel approx}), we can rewrite \eqref{eq:signal_model} as 
}
\begin{equation}
    r = \bof{w}^H (\bof{G}_y\otimes\bof{G}_z) \text{diag}(\bof{\theta})(\bof{H}_y\otimes\bof{H}_z) \bof{q}x + n.
    \label{eq:signal_model_kronecker_start}
\end{equation}
\al{To fully decouple the received signal as well as the optimization problem (\ref{eq:optimization_problem}) into horizontal and vertical components, we impose a  Kronecker product structure to the active and passive beamforming vectors (i.e., precoder, combiner, and IRS phase shifts) by defining
\begin{equation}
\bof{w}\doteq\bof{w}_y\otimes\bof{w}_z, \quad \bof{q}\doteq\bof{q}_y\otimes\bof{q}_z, \quad \bof{\theta}\doteq \bof{\theta}_y \otimes \bof{\theta}_z.
\label{eq:Kronecker beamforming}
\end{equation}
% \begin{equation}
%      r = \bof{w}^H \big[(\bof{G}_y\otimes\bof{G}_z) (\bof{\Theta}_y \otimes \bof{\Theta}_z ) (\bof{H}_y\otimes\bof{H}_z)\big] \bof{q} + n,
% \end{equation}
% which can be written as
% \begin{align}
%      &r = \bof{w}^H \big[(\bof{G}_y\bof{\Theta}_y \bof{H}_y)\otimes(\bof{G}_z\bof{\Theta}_z \bof{H}_z)\big] \bof{q} + n,
% \end{align}
% further assuming $\bof{q}$ and $\bof{w}$ factorization into horizontal and vertical domains as 
% \begin{align}
%     &r = (\bof{w}_y\otimes\bof{w}_z)^H (\bof{G}_y\bof{\Theta}_y \bof{H}_y)\otimes(\bof{G}_z\bof{\Theta}_z \bof{H}_z) (\bof{q}_y \otimes \bof{q}_z) + n,
% \end{align}
Using these definitions, we simplify (\ref{eq:signal_model_kronecker_start}) to
}
% \begin{align}
%     &r = \Scale[0.9]{(\bof{w}_y\otimes\bof{w}_z)^H(\bof{G}_y\otimes\bof{G}_z) (\bof{\Theta}_y \otimes \bof{\Theta}_z ) (\bof{H}_y\otimes\bof{H}_z) (\bof{q}_y \otimes \bof{q}_z) + n},
%     \label{eq:signal_model_kronecker_end}
% \end{align}
% and -- rearranging \eqref{eq:signal_model_kronecker_end} into horizontal and vertical domains -- we finally get:
\begin{align}
&r= s_y s_z+ n,\label{eq:factorized signal}
\end{align}
where $s_y\doteq \bof{w}_y^H\bof{G}_y \text{diag}(\bof{\theta}_y)\bof{H}_y\bof{q}_y$ is the horizontal domain desired signal component, and $s_z\doteq \bof{w}_z^H\bof{G}_z \text{diag}(\bof{\theta}_z)\bof{H}_z\bof{q}_z$ is the vertical domain desired signal component.
% \begin{align}
%    &r_y\doteq \bof{w}_y^H\bof{G}_y \text{diag}(\bof{\theta}_y)\bof{H}_y\bof{q}_y
%    \label{eq:factorized signal y}\\
%    &r_z\doteq \bof{w}_z^H\bof{G}_z \text{diag}(\bof{\theta}_z)\bof{H}_z\bof{q}_z.
%     \label{eq:factorized signal z}
% \end{align}
% It is now possible to separate the received signal in $r_y$ and $r_z$ and then estimate the active beamforming vectors at the \ac{BS} and the \ac{UE} and the passive beamforming at the \ac{IRS} for each component.
% The signal model described in eq.\ref{eq:rank 1 channel} represents a rank-one channel, i.e. \ac{LoS} link. For multi path channel we can impose the kronecker structure as follows
% \begin{align}
%      &\bof{H} = \sum^P_{i=1}   \underbrace{\alpha^{i}_H (\bof{h}_y^{i}_{\text{ar}}\bof{h}_y^{i}_{\text{dep}}^T)}_{\bof{H}_y^{i}} \otimes \underbrace{(\bof{h}_z^{i}_{\text{ar}}\bof{h}_z^{i}_{\text{dep}}^T)}_{\bof{H}_z^{i}} \\
%      &\bof{H} = \sum^P_{i=1} \bof{H}_y^{i}\otimes \bof{H}_z^{i}\approx  \Big(\sum^P_{i=1} \bof{H}_y^{i}\Big)\otimes \bof{H}_z,
% \end{align}
% then $\bof{H}_y$ will be equal to $\Big(\sum^P_{i=1} \bof{H}_y^{i}\Big)$ and now we can factorize the received signal in the same way as in eq.~\ref{eq:factorized signal}. 
\al{From \eqref{eq:factorized signal}, the \ac{SE} expression can then be written as 
% \begin{align}
%     &\text{SE} = \log_2\Big(1 + \frac{P|\left( \bof{w}_y^H\bof{G}_y\bof{\Theta}_y \bof{H}_y\bof{q}_y \right)\left(\bof{w}_z^H\bof{G}_z\bof{\Theta}_z \bof{H}_z\bof{q}_z\right)|^2}{\sigma_n^2} \Big)
% \end{align}
% which can be further written in compact as 
\begin{align}
    &\text{SE} = \log_2 \Big(1 + \frac{|s_y s_z|^2}{\sigma_n^2} \Big)=\log_2 \Big(1 + \text{SNR}_y \text{SNR}_z\Big), \label{snr}
\end{align}
where we have defined \textcolor{black}{$\text{SNR}_y = s_y s_y^*/\sqrt{\sigma^2}$ and $\text{SNR}_z = s_z s_z^*/\sqrt{\sigma^2}$}, as the \textit{horizontal} and \textit{vertical} SNRs, respectively.
Although imposing a Kronecker structure on these beamforming vectors restricts the solution space, a significant complexity reduction is achieved compared to the conventional ``full'' design, especially for large IRS panels. Exploiting the decoupled signal model in (\ref{eq:factorized signal}) two solutions are now derived to find the two-dimensional beamforming sets
$(\bof{w}_y,\bof{w}_z)$, $(\bof{q}_y,\bof{q}_z)$, and $(\bof{\theta}_y,\bof{\theta}_z)$.
}

\vspace{1ex}
\subsection{Kronecker Factorization (KF) Method}
\al{From the factorization of the received signal in (\ref{eq:factorized signal}), the full optimization problem \eqref{eq:optimization_problem} can be replaced by two smaller optimization sub-problems \fa{along each domain}, i.e.,
\begin{align}
    &\max_{\bof{w}_y,\bof{q}_y,\bof{\theta}_y} |(\bof{w}_y^H \bof{G}_y \text{diag}(\bof{\theta}_y)  \bof{H}_y \bof{q}_y)|^2
    \label{eq:optimization_problem_y}\\
    &\text{s.t.} \ ||\bof{w}_y|| = ||\bof{q}_y|| = 1 \ \text{and} \  \theta_{y} \in [-\pi,\pi],
   \nonumber % \label{eq:constrains_y}
\end{align}
\begin{align}
    &\max_{\bof{w}_z,\bof{q}_z,\bof{\theta}_z} |(\bof{w}_z^H \bof{G}_z \text{diag}(\bof{\theta}_z) \bof{H}_z \bof{q}_z)|^2
    \label{eq:optimization_problem_z}\\
    &\text{s.t.} \ ||\bof{w}_z|| = ||\bof{q}_z|| = 1 \ \text{and} \  \theta_{z} \in [-\pi,\pi].
    \nonumber %\label{eq:constrains_z}
\end{align}
\fa{individually maximizing the SNR in the $y$ and $z$ domains implies maximizing the overall SNR \eqref{snr}.}
% \ys{As shown in \eqref{eq:factorized signal}, the received signal $r_y$ and $r_z$ are independent to each other and are combined to obtain $r$. In this case, maximizing $r$ is the same as, independently, maximizing $r_y$ and $r_z$.}
% The optimization problem along the horizontal domain can be written as:
The solution to the individual problems \eqref{eq:optimization_problem_y} and \eqref{eq:optimization_problem_z} can be obtained from any state-of-the-art method. In this work, we consider the method of \cite{zappone_overhead_aware} but applied in each channel domain.
% we apply multiple rank-one \acp{SVD} in the channel components along horizontal and vertical domains to design the combiner, precoder and \ac{IRS} phase shifts along each domain independently. 
The steps of the \ac{KF} solution are detailed in Algorithm \ref{alg:SVD fac}.
}

% The same idea can be applied for the factorized received signal in equation \ref{eq:factorized signal}. A \ac{SVD} is applied to obtain a rank-one approximation of each channel component and the estimated singular vector are used to optimize as shown in algorithm \ref{alg:SVD fac}.

\begin{algorithm}[!t]
\SetAlgoLined
\nl Compute the truncated SVD of $\bof{G}_y$, $\bof{G}_z$, $\bof{H}_y$ and $\bof{H}_z$.\\
$\lambda_{\text{g}_y}\bof{u}_{\text{g}_y}\bof{v}_{\text{g}_y}^H\approx\bof{G}_y$, $\lambda_{\text{g}_z}\bof{u}_{\text{g}_z}\bof{v}_{\text{g}_z}^H\approx\bof{G}_z$, \\
    $\lambda_{\text{h}_y}\bof{u}_{\text{h}_y}\bof{v}_{\text{h}_y}^H\approx \bof{H}_y$ and $\lambda_{\text{h}_z}\bof{u}_{\text{h}_z}\bof{v}_{\text{h}_z}^H\approx \bof{H}_z$.\\
\nl Design the combiner as $\bof{w} = \bof{u}_{\text{g}_y}\otimes \bof{u}_{\text{g}_z}$.\\
\nl Design the precoder as $\bof{q} = \bof{v}_{\text{h}_y}\otimes \bof{v}_{\text{h}_z}$.\\
\nl Design the IRS phase shifts as\\ 
$\bof{\theta} = -\angle\{[\bof{v}_{\text{g}_y}^*\odot \bof{u}_{\text{h}_y}]\otimes[\bof{v}_{\text{g}_z}\odot\bof{u}_{\text{h}_z}]\}$
\caption{\label{alg:SVD fac}
Kronecker factorization (KF) method}
\end{algorithm} 

\vspace{-2ex}
\subsection{Third-Order Tensor (TOT) Method}
\al{The second method exploits the horizontal and vertical factorizations of the received signal from a tensor modeling perspective. Starting from (\ref{eq:factorized signal}), let us rewrite $s_y$ and $s_z$ components of the received signal as
\begin{align}
    &s_y=\bof{w}_y^H \bof{G}_y \text{diag}(\bof{\theta}_y)\bof{H}_y \bof{q}_y=(\bof{q}_y^T\otimes \bof{w}_y^H)\bof{F}_y\bof{\theta}_y, \nonumber \\
    &s_z=\bof{w}_z^H \bof{G}_z \text{diag}(\bof{\theta}_z)\bof{H}_z \bof{q}_z=(\bof{q}_z^T\otimes \bof{w}_z^H)\bof{F}_z\bof{\theta}_z,\nonumber
    %\text{vec}(\bof{G}_y\bof{\Theta}_y\bof{H}_y),
\end{align}
where we have defined $\bof{F}_y \doteq \bof{H}^T_y \diamond \bof{G}_y \in \mathbb{C}^{K_yM_y \times N_y}$ and $\bof{F}_z \doteq \bof{H}^T_z \diamond \bof{G}_z \in \mathbb{C}^{K_zM_z \times N_z}$ as the combined $y$ and $z$ domains Khatri-Rao channels, respectively.
%and used the property $\text{vec}(\bof{G}_y\text{diag}(\bof{\theta}_y) \bof{H}_y)=(\bof{H}_y^T\diamond \bof{G}_y)\bof{\theta}_y=\bof{F}_y\bof{\theta}_y$.
}
% \begin{align}
%     \text{vec}(\bof{G}_y \bof{\Theta}_y \bof{H}_y) = (\bof{H}_y^T\diamond \bof{G}_y)\bof{\theta}_y,
% \end{align}
%Let us define $\bof{F}_y \doteq \bof{H}^T_y \diamond \bof{G}_y \in \mathbb{C}^{K_yM_y \times N_y}$ as the combined $y$-domain Khatri-Rao channel. 
\al{We can rearrange the elements of $\bof{F}_y$ and $\bof{F}_z$ in third-order tensors $\mathcal{F}_y \in \mathbb{C}^{K_y \times M_y \times N_y}$ and $\mathcal{F}_z \in \mathbb{C}^{K_z \times M_z \times N_z}$ \textit{via} the mappings  $[\mathcal{F}_y]_{k_y,m_y,n_y} \doteq [\bof{F}_y]_{(k_y-1)M_y+m_y,n_y}$ and $[\mathcal{F}_z]_{k_z,m_z,n_z} \doteq [\bof{F}_z]_{(k_z-1)M_z+m_z,n_z}$,
where $k_t=1, \ldots, K_t$, $m_t=1,\ldots, M_t$, $n_t=1, \ldots, N_t$, $t \in \{y,z\}$.
The set of active and passive beamforming vectors that individually maximize the horizontal ($y$-domain) and vertical ($z$-domain) SNRs can be found by independently solving the following problems
\begin{align}
    &\max_{\bof{w}_y,\bof{q}_y,\bof{\theta}_y} || \mathcal{F}_y \times_1 \bof{w}_y \times_2 \bof{q}_y \times_3 \bof{\theta}_y||^2\label{eq:third order opt prob y} \\
     &\text{s.t.} \ ||\bof{w}_y|| = ||\bof{q}_y|| = 1 \ \text{and} \  \theta_y \in [-\pi,\pi] \nonumber
\end{align}   
\begin{align}
    &\max_{\bof{w}_z,\bof{q}_z,\bof{\theta}_z} || \mathcal{F}_z \times_1 \bof{w}_z \times_2 \bof{q}_z \times_3 \bof{\theta}_z||^2\label{eq:third order opt prob z}\\
     &\text{s.t.} \ ||\bof{w}_z|| = ||\bof{q}_z|| = 1 \ \text{and} \  \theta_z \in [-\pi,\pi] \nonumber
\end{align}
}

%The tensor $\ten T$ is referred to here as \textit{composite channel tensor}, whose first, second, and third modes have dimensions corresponding to the number $M$ of transmit antennas, $L$ of receive antennas, and $N$ of IRS reflecting elements.
\al{These problems can be solved by applying the \ac{HOSVD} \cite{Kolda2009} to the tensors $\mathcal{F}_y$ and $\mathcal{F}_z$. The solutions to $\bof{w}_y$  $\bof{q}_y$, and $\bof{\theta}_y$ correspond respectively to the 1-mode, 2-mode, and 3-mode dominant left singular vectors of $\mathcal{F}_y$. Likewise, $\bof{w}_z$  $\bof{q}_z$, and $\bof{\theta}_z$ are found as the  1-mode, 2-mode, and 3-mode dominant left singular vectors of $\mathcal{F}_z$ (we refer the reader to \cite{Kolda2009} for further details on the HOSVD algorithm). The steps of the \ac{TOT} method are summarized in Algorithm \ref{alg:third order tensor}.
}

% Moreover, note that $\mathcal{F}_y$ is a rank-one tensor defined as
% \begin{align}
%     &\mathcal{F}_y = \bof{c}_y(\mu_\text{ue}) \circ \bof{a}_y(\mu_\text{bs})\circ \bof{f}_y(\mu_{\text{irs}_A},\mu_{\text{irs}_D}),
% \end{align}

% Similarly, for the vertical domain, we can define a third-order tensor $\mathcal{F}_z$ as
% \begin{align}
%     &\mathcal{F}_z = \bof{c}_z(\mu_\text{ue}) \circ \bof{a}_z(\mu_\text{bs})\circ \bof{f}_z,
% \end{align}
% where $\mathcal{F}_z \in \mathbb{C}^{K_z \times M_z \times N_z}$, and -- analogously -- the vertical tensor optimization problem can be written as
% \begin{align}
%      &\max_{\bof{w}_z,\bof{q}_z,\bof{\theta}_z} || \mathcal{F}_z \times_1 \bof{w}_z \times_2 \bof{q}_z \times_3 \bof{\theta}_z||^2\\
%      &\text{s.t.} \ ||\bof{w}_z|| = ||\bof{q}_z|| = 1 \ \text{and} \  \theta_z \in [-\pi,\pi].
%     \label{eq:third order opt prob z}
% \end{align}

% \begin{align}
%     [\mathcal{F}_y]_n = \bof{B}^{(n)}\bof{F}^{(n)}\bof{T}^{(n)}
% \end{align}
% \begin{align}
%      [\mathcal{F}_z]_n = \bof{C}^{(n)}\bof{L}^{(n)}\bof{O}^{(n)}
% \end{align}
\begin{algorithm}[!t]
\SetAlgoLined
\nl Compute the \ac{HOSVD} of $\mathcal{F}_y$ and $\mathcal{F}_z$ to obtain \\
$\{\bof{w}_y, \bof{q}_y, \bof{\theta}_y\}$ and $\{\bof{w}_z, \bof{q}_z, \bof{\theta}_z\}$, respectively.\\
\nl Design the combiner as $\bof{w} = \bof{w}_y \otimes \bof{w}_z$.\\
\nl Design the precoder as $\bof{q} = \bof{q}^*_y \otimes \bof{q}^*_z$.\\
\nl Design the IRS phase shifts as $\bof{\theta} = -\angle[\bof{\theta}_y \otimes \bof{\theta}_z]$.
\caption{\label{alg:third order tensor}
Third-order tensor (TOT) method}
\end{algorithm} 
\vspace{-0.1in}
\subsection{Complexity Analysis}
\al{Both KF and TOT methods rely on (truncated SVD-based) rank-one approximation procedures. Note that a single low-rank matrix approximation has a complexity } \ys{$O(ijv)$ \cite{truncated_svd}}, where $i$ is the number of rows, $j$ is the number of columns and $v$ is the rank. For the rank-one approximation, this complexity is reduced to \ys{$O(ij)$}.
\ys{In the baseline solution of \cite{zappone_overhead_aware}, the precoder, combiner, and IRS phase shifts} \al{are determined from the dominant left singular vectors of the full channel matrices $\bof{H}$ and $\bof{G}$, resulting in $O(N(M+K))$.} 
\al{The \ac{KF} method solves the beamforming problem for each channel dimension separately, which implies solving four rank-one matrix approximation procedures to $\bof{G}_y$, $\bof{G}_z$, $\bof{H}_y$ and $\bof{H}_z$.
Thus, the overall complexities of the horizontal and vertical problems are  respectively given as $O~(N_y(M_y~ + ~K_y))$ and $O(N_z(M_z + K_z))$. Finally, the complexity of the \ac{TOT} method corresponds to that of computing two independent HOSVDs for the horizontal and vertical optimization problems, where the individual HOSVDs have complexities $O(3K_y M_y N_y)$ and $O(3K_z M_z N_z)$, respectively. The complexities are summarized  in Table~\ref{tab:computational cost}.
}
\vspace{-0.1in}
\section{Simulation Results} 
	\label{sec:simulations}
We consider a \ac{MIMO} communication system where the \ac{BS} is equipped with $M = 128$ antenna elements, composed of $M_y = 16$ and $M_z = 8$ along horizontal and vertical domains, serves a single user with $K = 16$ \textcolor{black}{antenna} elements composed of $K_y = K_z = 4$ antenna elements along horizontal and vertical domains, respectively. The communication is assisted by an \ac{IRS} equipped with $N=100$ reflecting elements composed of $N_y = N_z = 10$ reflecting elements along horizontal and vertical domains, respectively. The \ac{EoA} and the \ac{EoD} are generated from the uniform distribution $\theta_{\text{bs}},\theta_{\text{irs}_A},\theta_{\text{irs}_D},\theta_\text{ue}\sim\text{U}~[90^\circ - \delta,90^\circ + \delta]$, where $\delta$ is the elevation spread. Also, the \ac{AoA} and the \ac{AoD} are generated from the uniform distribution $\phi_{\text{ue}},\phi_{\text{irs}_A},\phi_{\text{irs}_D},\phi_\text{ue} \sim\text{U}~[-60^\circ,60^\circ]$. Also, $4$ paths are considered for $\bof{G}$ and $\bof{H}$.  
For comparison, the \al{SVD-based joint active and passive beamforming algorithm} proposed in \cite{zappone_overhead_aware} is used as \al{the} baseline solution \al{serving as a reference for comparison}.

\begin{table}[!t]
    \caption{Computational complexity}
    \centering
    \begin{tabular}{|c|c|}
    \hline
            Solution & Complexity \\ \hline
          Baseline \cite{zappone_overhead_aware} &$ O(N(M+K)) $\\\hline
          KF & $O(N_y(M_y + K_y) + N_z(M_z + K_z))$\\\hline
          % SOT & \thead{$\Scale[0.8]{O(K_z + K_y M N) + O(K_y + K_z M N) + O(M_z + M_y K N) +}$\\ $\Scale[0.8]{O(M_y + M_z K N) + O(N_z + N_y K M) + O(N_y + N_z K M)}$}\\\hline
          TOT & $O(3(K_y M_y N_y + K_z M_z N_z)) $\\ \hline
          
          % HPS & $O(N)$\\\hline
          % GSS & $\Scale[0.9]{O(M + N) + O(N + K) + O(N) + 4O(TL)}$\\\hline
    \end{tabular}
    \label{tab:computational cost}
\end{table}

\begin{figure}[!t]
    \centering
    \includegraphics[scale = 0.2]{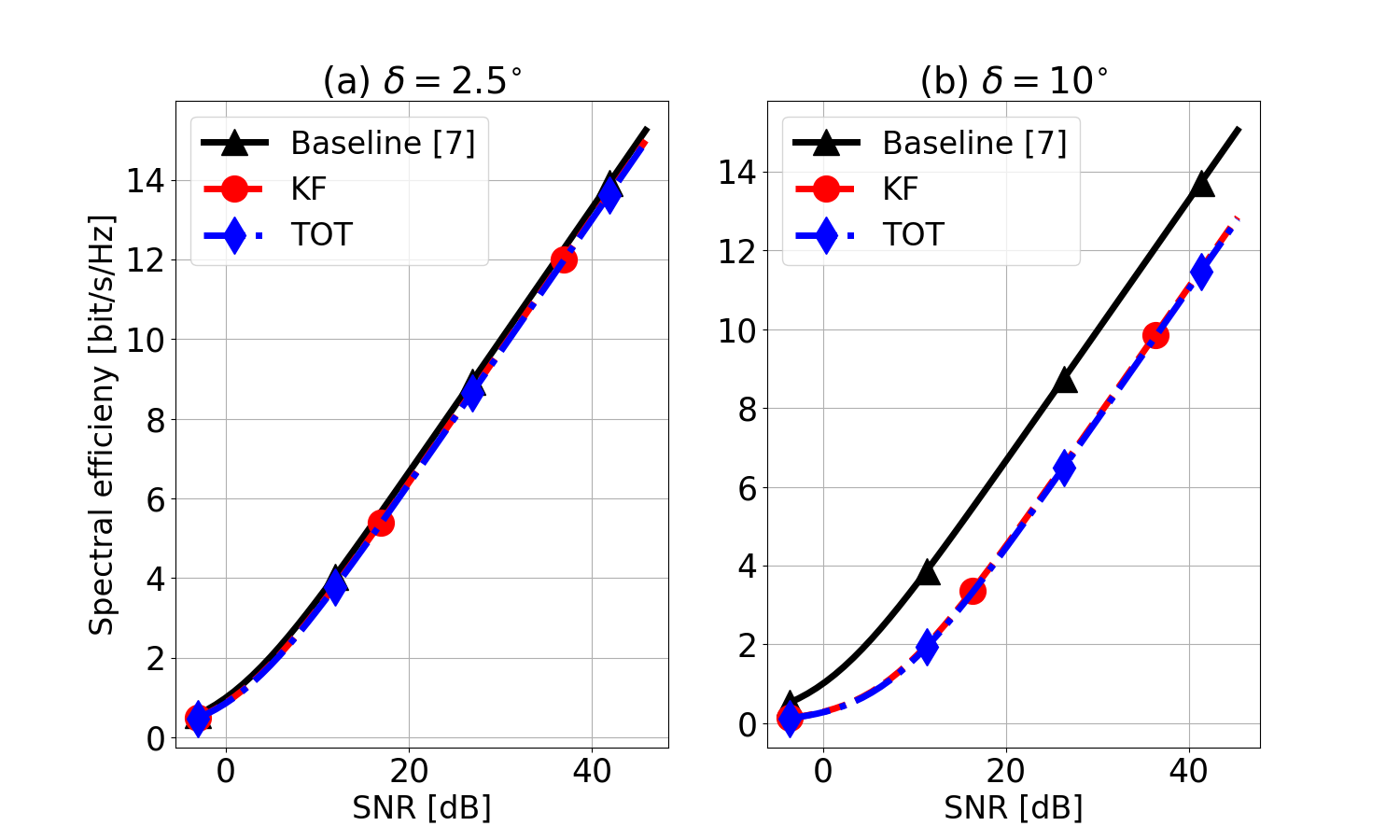}
    \caption{\ys{SE of the proposed methods compared to the baseline \cite{zappone_overhead_aware} for $\delta = 2.5^\circ$ and $\delta = 10^\circ$, varying the SNR.}}
    \label{fig:SE vs SNR}
\end{figure}
% \begin{figure}[!t]
%     \centering
%     \includegraphics[scale = 0.3]{figs/SE vs spread tensor.png}
%     \caption{\ys{SE of the proposed methods in comparison with the baseline \cite{zappone_overhead_aware} varying the $\delta$ for a fixed SNR.}}
%     \label{fig:SE vs delta}
% \end{figure}

Fig. \ref{fig:SE vs SNR} shows the performance of the proposed algorithms varying the SNR for different elevation spread values. \textcolor{black}{Note that the approximation error associated with the separable channel structure assumption in \eqref{eq:H channel approx} and \eqref{eq:G channel approx} increases as a function of the elevation spread}. The results show that the proposed KF and TOT solutions have a slight performance degradation compared with the baseline \al{method of} \cite{zappone_overhead_aware}. For example, in Fig 1a, the performance degradation is $0.3$ bit/s/Hz, while in Fig 1b, it is equal to $2.3$ bit/s/Hz. \al{Indeed}, \ys{since the optimizations problems in \eqref{eq:third order opt prob y} and \eqref{eq:third order opt prob z} are the same as the ones in \eqref{eq:optimization_problem_y} and \eqref{eq:optimization_problem_z}, respectively, both solutions have the same \ac{SE}}. \textcolor{black}{It is important to highlight that, in scenarios where the approximations in \eqref{eq:H channel approx} and \eqref{eq:G channel approx} become exact, the imposed Kronecker structure of the passive and active beamforming vectors in \eqref{eq:Kronecker beamforming} does not degrade the SE, i.e., the proposed KF and TOT solutions achieve the same SE as the baseline solution \cite{zappone_overhead_aware}}.
\begin{figure}[!t]
    \centering
    \includegraphics[scale = 0.2]{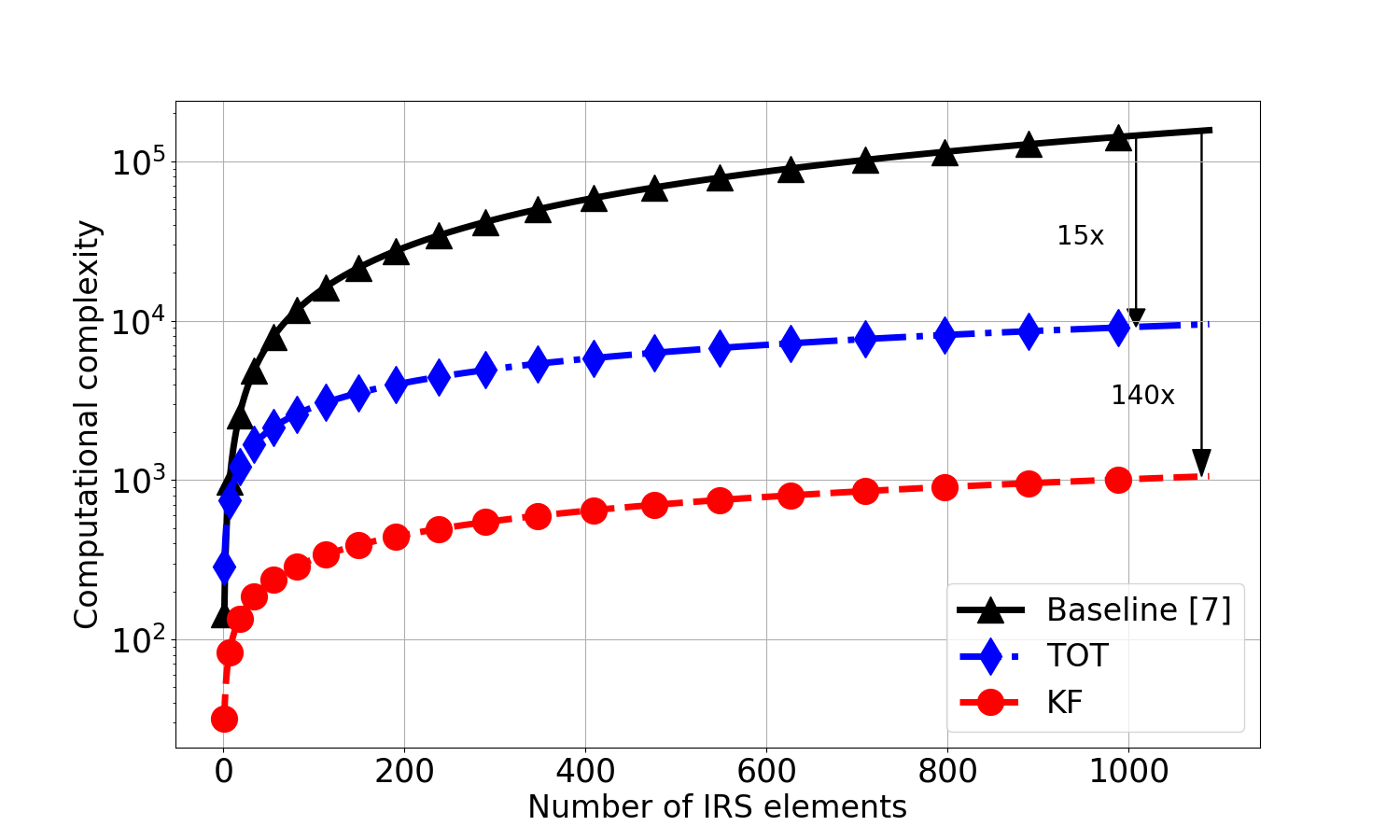}
    \caption{\ys{Computational complexity of the proposed solutions and the baseline \cite{zappone_overhead_aware} for fixed $M$ and $K$ while varying $N$.}}
    \label{fig:complexity curves}
\end{figure}

Fig. \ref{fig:complexity curves} shows the computational complexity as a function of the number of IRS reflecting elements. By exploiting the Kronecker factorization structure of the channels, the KF and TOT solutions have lower slopes when compared to the baseline \al{algorithm \cite{zappone_overhead_aware}.} Although KF and TOT have small degradation in performance, they can significantly reduce the computational complexity when moderate or large numbers of IRS elements are considered and, as shown previously, depending on the scenario they can have very close performance compared to the baseline solution. For example, \al{for} $1000$ reflecting elements the baseline solution \cite{zappone_overhead_aware} is \al{$140$ and $15$ times more complex than the proposed KF and TOT solutions, respectively.}

\begin{figure}[!t]
    \centering
    \includegraphics[scale = 0.2]{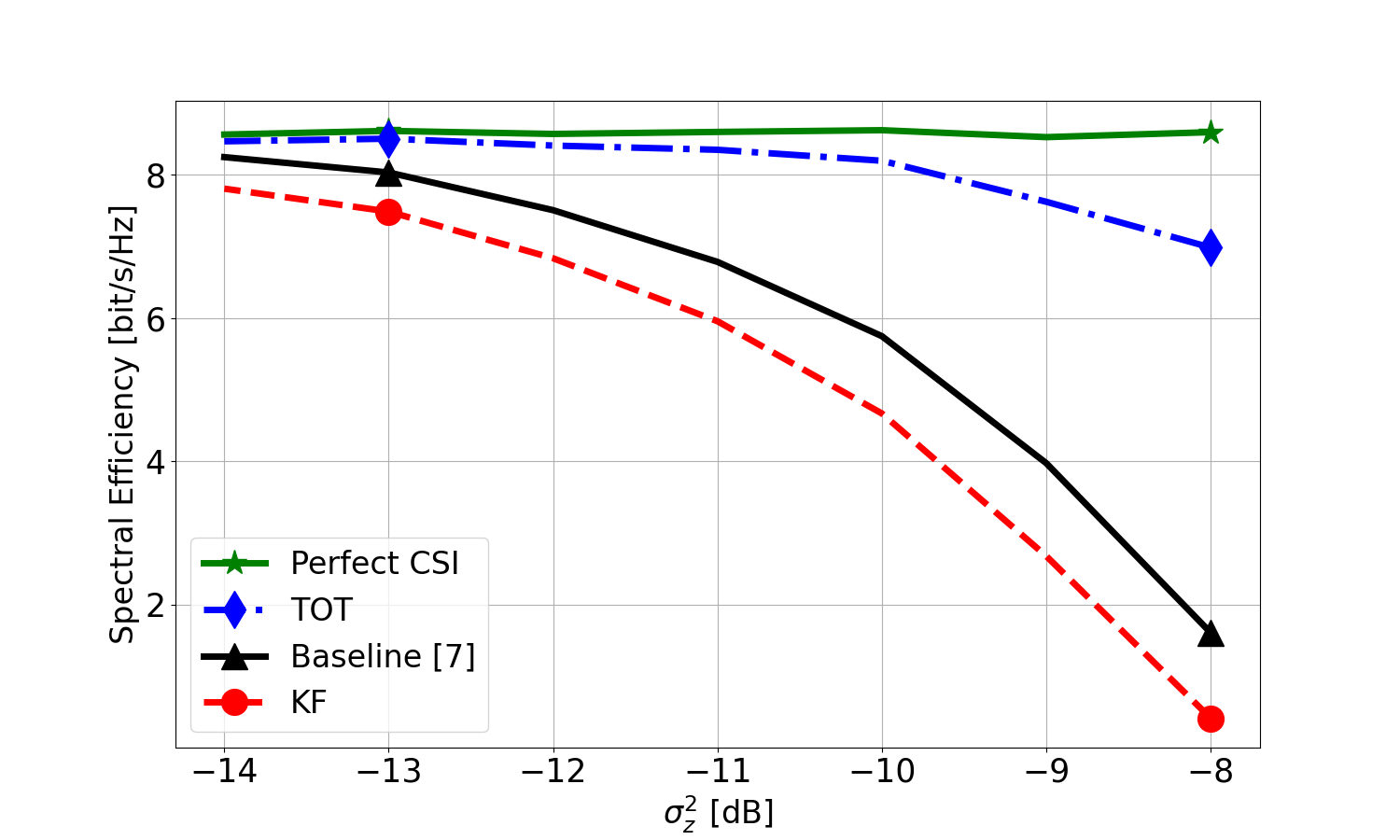}
    \caption{\ys{SE of the proposed methods and the baseline method of \cite{zappone_overhead_aware} \al{under} imperfect CSI.}}
    \label{fig:SE noisy}
\end{figure}

Figure \ref{fig:SE noisy} depicts the SE of the proposed algorithms \al{when channel estimation error is taken into account, \textcolor{black}{considering $\text{SNR} = 20$dB}. To this end, let $\bof{\hat{F}} \doteq \bof{H}^T\diamond \bof{G} + \bof{Z}$ be the noisy estimate of the combined (Khatri-Rao) channel obtained after LS estimation and matched filtering with the known orthogonal pilots and optimal DFT IRS phase shift matrices (details can be found in \cite{Jensen20}),} \fa{where $\bof{Z} \sim \mathcal{C}\mathcal{N}(0,\sigma_z^2)$ \al{accounts for} the channel estimation error.} \al{Also, define $\bof{\hat{F}}_y$ and $\bof{\hat{F}}_z$ as the noisy estimates of the $y$ and $z$ channel components of $\bof{\hat{F}}$ (obtained according to \cite{asim2023}). The baseline method of \cite{zappone_overhead_aware} optimizes the beamforming vectors from the estimates of $\bof{H}$ and $\bof{G}$ obtained after a Khatri-Rao factorization step applied to the noisy estimate $\bof{\hat{F}}$. The KF method relies on the decoupled estimates of $\bof{\hat{H}}_y,\bof{\hat{H}}_z,\bof{\hat{G}}_y$,  $\bof{\hat{G}}_z$ obtained after Khatri-Rao and Kronecker factorization steps, while the TOT method operates directly on the estimates $\bof{\hat{F}}_y$ and $\bof{\hat{F}}_z$, avoiding the Khatri-Rao factorization step.}
\yss{The perfect CSI curve \al{corresponds to} the baseline method \cite{zappone_overhead_aware} assuming perfect CSI knowledge}. \al{As shown in} Fig. \ref{fig:SE noisy}, for  higher \al{values of} $\sigma_z^2$ TOT \al{outperforms} the baseline and the KF methods due to the processing required to obtain $\bof{\hat{F}}_y$ and $\bof{\hat{F}}_z$ \fa{which rejects part of the noise}. The KF has \al{the worst} performance due to the \al{scaling} ambiguities \fa{in the Khatri-Rao factorization step that propagates to the subsequent stage where the horizontal and vertical channel matrices are estimated}. \al{Hence, although TOT is more computationally complex than KF, under moderate channel estimation error, it has a better SE performance. For example, assuming} $\sigma_z^2 = -8$dB the TOT, baseline, and KF solutions \al{present a SE gap of $1.6$, $7$, and $8.2$ bits/s/Hz compared to the perfect CSI curve. These results show the involved tradeoffs between complexity and performance under imperfect CSI.}
\vspace{-0.1in}
\section{Conclusion}
\vspace{-0.05in}
\label{sec:conclusion}
We proposed \al{two low-complexity joint active and passive beamforming designs for IRS-assisted MIMO systems. Our methods exploit} the geometrical channel structure to split the \al{involved optimization problems} into vertical and horizontal domains. The proposed KF and TOT algorithms significantly reduce the computational complexity with negligible SE degradation \al{ compared to the reference method under perfect CSI, while the TOT method provides the best SE performance when channel estimation noise is taken into account} in the proposed optimization \al{due to more efficient noise rejection}.}

% \appendix

% \section{A}
%  Noting that $\bof{H}_y$ and $\bof{G}_y$ are rank-one matrices, we have
% \begin{align}
%      &\bof{F}_y = \alpha_y[\bof{a}_y(\mu_\text{bs})\bof{b}_y^T(\mu_{\text{irs}_A})] \diamond [\bof{c}_y(\mu_\text{ue})\bof{d}_y^T(\mu_{\text{irs}_D})],
% \end{align}
% where $\alpha_y$ is a scaling factor. Using the property $\bof{AB}\diamond\bof{CD} = (\bof{A}\otimes\bof{C})(\bof{B}\diamond\bof{D})$, we obtain
% \begin{align}
%     &\bof{F}_y = [\bof{a}_y(\mu_\text{bs}) \otimes \bof{c}_y(\mu_\text{ue})][\bof{b}_y^T(\mu_{\text{irs}_A}) \diamond \bof{d}_y^T(\mu_{\text{irs}_D})],
% \end{align}
% or, compactly,
% \begin{align}
%     \bof{F}_y = [\bof{a}_y(\mu_\text{bs}) \otimes \bof{c}_y(\mu_\text{ue})]\bof{f}_y^T(\mu_{\text{irs}_A},\mu_{\text{irs}_D}),
%     \label{eq:Uy matrix}
% \end{align}
% where $\bof{f}_y(\mu_{\text{irs}_A},\mu_{\text{irs}_D})  \doteq \bof{b}_y^T(\mu_{\text{irs}_A}) \diamond \bof{d}_y^T(\mu_{\text{irs}_D})$.
% \al{We can rearrange the elements of $\bof{F}_y$ in third-order tensor $\mathcal{F}_y \in \mathbb{C}^{K_y \times M_y \times N_y}$ \textit{via} the the mapping  $[\mathcal{F}_y]_{k,m,n} \doteq [\bof{F}_y]_{(k-1)M+m,n}$,
% where $k=1, \ldots, K_y$, $m=1,\ldots, M_y$, $n=1, \ldots, N_y$. 

\bibliographystyle{IEEEtran} 
\bibliography{Ref2020}

\end{document}